\documentclass[twocolumn,prl]{revtex4}

\ifx\pdfoutput\undefined
  \usepackage[dvips]{graphicx}
\else
  \usepackage[pdftex]{graphicx}
\fi

\usepackage{dcolumn}
\usepackage{amsmath}
\usepackage{latexsym}

\begin{document}

\title[Short Title]{Cooper pair cotunneling in single charge
transistors with dissipative electromagnetic environment}% Force line breaks with \\

\author{S.~V.~Lotkhov, S.~A.~Bogoslovsky, A.~B.~Zorin and J.~Niemeyer}
\affiliation{Physikalisch-Technische Bundesanstalt, Bundesallee
100, 38116 Braunschweig, Germany}

\date{\today}% It is always \today, today, but you may specify any date with \date.

\begin{abstract}
We observed current-voltage characteristics of superconducting
single charge transistors with on-chip resistors of $R\sim
R_Q=h/4e^2\approx 6.45\,$k$\Omega$, which are explained in terms
of Cooper-pair cotunneling. Both the effective strength of
Josephson coupling and the cotunneling current are modulated by
the gate-induced charge on the transistor island. For increasing
values of the resistance $R$ we found the Cooper pair current at
small transport voltages to be dramatically suppressed.

%\verb+\pacs{#1}+ command.
\verb  PACS numbers: 74.50.+r, 74.25.Fy, 74.40.+k
\end{abstract}

\maketitle

During the last decade superconducting circuits with mesoscopic
tunnel junctions have attracted much attention of researchers.
Remarkable quantum properties of these circuits, which stem from
interaction between Josephson and charging effects, enable the
construction of charge-based qubits \cite{qubit}, sensitive
electrometers \cite{CP-electr}, amplifiers \cite{BOT}, etc. Since
the Josephson supercurrent itself is of an elastic nature,
tunneling of Cooper pairs in such circuits is dramatically
influenced by the Coulomb interaction and dissipation.

The tunneling in the two-junction system (single Cooper pair
transistor) with a dissipative environment was recently
theoretically studied by Wilhelm et~al. \cite{Wilhelm} and
experimentally investigated by Kycia et al. \cite{Kycia} and Lu
et~al. \cite{Lu} who used the advantage of locally tunable
dissipation associated with the conductance of 2DEG beneath the
transistor structure. These papers were focused on the regime of
sequential tunneling of Cooper pairs (STCP) dominating when the
electric potential of the transistor island was tuned "in
resonance" (see parameter diagram in Fig.~3 of
Ref.~\cite{Wilhelm}). Although a quantum regime of simultaneous
pair tunneling across both junctions, i.e. cotunneling of Cooper
pairs (CTCP), was roughly evaluated in Ref.~\cite{Wilhelm}, it
had not been, to our knowledge, studied experimentally.

Under the term "cotunneling" we understand discrete tunneling of
$single$ Cooper pairs across a whole transistor \cite{cotunn}.
Although numerous experiments show that Cooper pair transistors
without deliberately attached resistors exhibit either a
gate-dependent switching current at ramp-current bias or a
current peak at low voltage bias (see, e.g., Ref.\cite{Amar}),
these data indicate to a through supercurrent. Due to unavoidable
dissipation leading to fluctuations of the overall Josephson phase
$\phi$, the observable value of the supercurrent is non-zero at
finite transport voltage. Phase $\phi$ is still a good variable,
while the number of transferred Cooper pairs is meaningless, so
the term "single" is not applicable to this regime.

When an appreciable impedance is introduced into the network,
i.e. ${\rm Re}Z =R \sim R_Q =h/4e^2\approx 6.45\,$k$\Omega$,
quantum fluctuations of phase $\phi$ increase dramatically, and
this leads to incoherent tunneling of individual pairs across the
transistor, i.e. CTCP. In this paper we investigate the CTCP
effect in Al transistors with thin-film Cr local resistors of
different values $R$ inserted into the bias leads. We observe the
gate and bias voltage dependencies of CTCP current and
demonstrate that a significant suppression of CTCP at small bias
voltage can be achieved by a further increase in the resistance
$R$.

The Cooper pair transistors were fabricated using the angle
evaporation technique for three metal shadows
(Cr/Al-1(oxidized)/Al-2) evaporated at different angles through
the same mask. Details of the fabrication process can be found
elsewhere \cite{Napoli}. We put on the same chip a number of
nominally identical symmetrical transistors (with tunnel junctions
of type Al/AlO$_{\rm x}$/Al with dimensions 40 nm $\times$ 80 nm)
differing only in the lengths of the Cr microstrips (each
0.1~$\mu$m - wide and 7~nm - thick), ranging from 0.32~$\mu$m to
3~$\mu$m. The specific film resistivity was from 5.5--7 k$\Omega$
per $\mu$m of length yielding values of resistance $R$ in the
range of 2--20~k$\Omega$.

The measurements were performed in the dilution refrigerator at
the base temperature of about 20~mK. The sample was mounted in a
copper shielding box. All bias and gate lines were supplied with
microwave filters, made of Thermocoax$^{\rm\textregistered}$
cable located in the coldest part of the setup. The electric
diagram of the sample bias is shown in the upper inset on Fig.~1.
The normal-state tunnel resistance of the transistors, $R_{\rm
tr}=R_1+R_2$, where $R_{1,2}$ are tunnel resistances of
individual junctions, was remarkably uniform and ranged from
40~k$\Omega$ to 44~k$\Omega$. The superconducting gap and the
corresponding values of the Ambegaokar-Baratoff critical current
of the junctions (assuming symmetry of the transistors) were
estimated to be $\Delta_{\rm Al}\approx 200~\mu$eV and
$I_c^{0}\approx 16$~nA, respectively. These data yield a
characteristic Josephson energy of junctions
$E_{J1,J2}~=~E_{J}\equiv \frac{\Phi_0}{2\pi}I_c^{0} ~\approx~$ 30
$\mu$eV, where $\Phi_0=\frac{h}{2e}$ is the flux quantum. From
the normal-state $I-V$ curves we evaluated a characteristic
charging energy of the transistor island $E_{c}\equiv
e^2/2C_{\Sigma}\approx$ 150--160~$\mu$eV, where $C_{\rm {\Sigma}}
= C_{1}+C_{2}+C_{g} \approx 530$~aF (typical value) is the total
capacitance of the island and $C_{1,2}=C~\approx 260$~aF and
$C_g\approx 9$~aF $\ll C$ are the capacitances of junctions and
gate, respectively. Due to the small $C_g$, the gate-induced
charge is fixed, $Q_0=C_g V_g+Q_{00}$, where $Q_{00}$ is the
background charge. The reasonably small ratio of characteristic
energies in our samples, $\lambda_0=E_{J}/E_{c}\approx 0.2$,
allowed us both clear observation and transparent interpretation
of the CTCP effect ($\propto \lambda_0^4$). Moreover, linearity
and small resistor lengths let us consider them as lumped
elements (pure resistors) and neglect in our model their
distributed self-capacitance ($\approx$ 60~aF/$\mu$m).

$I-V$ characteristics of superconducting single charge
transistors are generally featured by several contributions to the
net current. These contributions result from the tunneling of
Cooper pairs, single quasiparticles and combining tunneling of
both pairs and quasiparticles. At small transport voltage and
temperature ($eV, k_BT \ll 2E_c, 2\Delta_{\rm Al}$), the
tunneling of quasiparticles cannot make a notable contribution to
the net current because of the exponentially large subgap
resistance. In this limit, the Cooper pair (Josephson) current
component is dominant. Still, the rare events of quasiparticle
tunneling (poisoning of the island) at $|Q_0| \geq e/2$ were
possible and they resulted in 1e-periodic gate dependence of the
$I-V$ curves (see bottom inset in Fig.~1). Therefore, slow
ramping of the gate voltage (when measuring gate-modulation
curves) provided cyclic variations of the polarization charge on
the island $Q_0$ in the range from $-e/2$ to $e/2$. Within this
range of $Q_0$, change of the island charge by 2e would require
an energy $\delta E_{\rm ch}$ ranging between $2E_c$ (at $Q_0=\pm
e/2$) and $4E_c$ (at $Q_0=0$). Therefore, the process of Cooper
pair transfer with "landing" on the island, i.e. the STCP, was
exponentially suppressed, $\propto \exp (-\delta E_{\rm
ch}/k_BT)$, so only the CTCP contributed to the net current.

Figure~1 shows the $I-V$ curves of three samples with resistance
$R$ of nominally zero value (the actual impedance in this case is
of the order of the vacuum impedance $Z_0 \approx 377\,\Omega$,
although its frequency dependence is unknown), 3~k$\Omega$ and
5~k$\Omega$. One can see a remarkable change from the peak (cf.
Fig.~1 in Ref.~\cite{Amar}) to step-like and then to an almost
linear shape of the $I-V$ curve that can be interpreted as a
growth of fluctuations of the total Josephson phase leading to a
depression of the supercurrent feature.

\begin{figure}[t]
\begin{center}
\leavevmode
\includegraphics[width=3.0in]{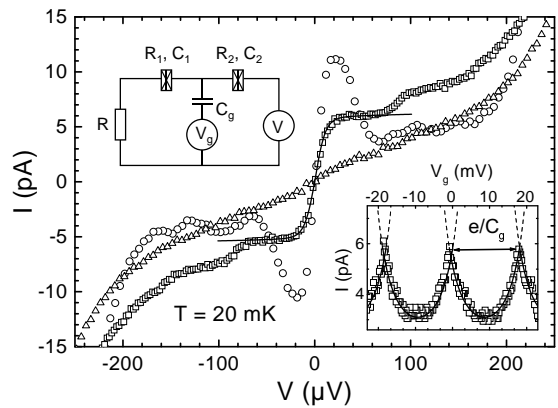}
\caption {The electric circuit diagram (upper inset) and $I-V$
curves of three Cooper pair transistors with different nominal
values of the resistance $R$ (circles - 0, squares - 3~k$\Omega$
and triangles - 5~k$\Omega$) at gate voltage values $V_g$
corresponding to maximum of current. The solid line shows computed
data assuming a symmetric transistor with $E_J=26~\mu$eV and a
resistor of $R=3.3$~k$\Omega$ with an effective temperature of
90~mK. The gate dependence in the transistor with $R=3$~k$\Omega$
at $V=30~\mu$V is presented in the bottom inset, where the solid
(dashed) line shows the data calculated within an experimentally
accessible (inaccessible) range.} \label{Scheme}
\end{center}
\end{figure}

A similar behavior of a single junction was described by Averin
et al. \cite{AvNazOd} in terms of incoherent tunneling of pairs
yielding current $I=2e[\Gamma(2eV)-\Gamma(-2eV)]$ and a tunneling
rate
\begin{equation}
\label{Gamma} \Gamma (E)=\frac{\pi}{2\hbar} E_{J}^2 P(E),
\end{equation}
where the function $P(E)$ with $E=2eV$ describes the probability
of energy dissipation in the environment network. The experimental
observation of this regime was reported by Kuzmin et al.
\cite{Kuzmin}.

In the case of a transistor, one may consider $E_J$ in
Eq.~(\ref{Gamma}) as an effective Josephson coupling dependent on
the gate voltage $V_g$. To support this idea, we observed the
shapes of the $I-V$ curves similar to those given by the
$P$-theory for single junctions with ohmic environment
\cite{AvNazOd,IngGrab}. Moreover, the observed shapes were
immutable to $V_g$ whose variation led only to scaling of the
current.

To model the behavior of our $I-V$ curves, we present the
Josephson coupling term of the transistor Hamiltonian, $H_{J}=-
E_{J1}\cos \varphi _1- E_{J2}\cos\varphi_2$, in the form
\begin{equation}
\label{HJos2} H_{J}(\phi,\varphi)  = - E_J^{\rm tr} (\phi )\cos
\left[ {\varphi + \gamma (\phi )} \right],
\end{equation}
where $E_J^{\rm tr}  = \left( {E_{J1}^2  + E_{J2}^2  + 2E_{J1}
E_{J2} \cos \phi } \right)^{1/2} $, $\tan \gamma = \left[ {\left(
{E_{J1}  - E_{J2} } \right)/\left( {E_{J1}  + E_{J2} } \right)}
\right]\tan (\phi /2)$ and $\phi=\varphi_1+\varphi_2$
($\varphi_{i}$ is the phase drop across the $i$-th junction). The
difference phase $\varphi=\frac{1}{2}(\varphi_1-\varphi_2)$ is a
variable conjugate to the island charge $Q=-2ei
\frac{\partial}{\partial\varphi}$. $Q$ enters the charging energy
term $H_{\rm ch}(Q)=E_c[(Q+Q_0)/e]^2$. For a given value of
operator $\phi$, the transistor Hamiltonian $H_{\rm tr}= H_{\rm
ch}(Q)+H_{J}(\phi,\varphi)$ is similar to that of a single
junction. Taking the fact into account that the commutator
$[\phi,H_{\rm tr}]=0$, we can diagonalize $H_{\rm
tr}(\phi,\varphi,Q)=H_{\rm tr}(\phi,q,n)$ in the basis of the
Bloch states $|q,n\rangle$, where $q$ is the quasicharge and
$n=0,1,2,...$ the band index \cite{LiZo}. Assuming that only the
ground state, $n=0$, is occupied (i.e., the effects of thermal,
photon-assisted and Zener breakthrough excitations of higher bands
are negligibly small) and noting that variable $q$ is decoupled
from environmental degrees of freedom due to small $C_g$ and
fixed by $V_g$, we present the transistor Hamiltonian as
\begin{equation}
\label{Htr1} H_{\rm tr}= E_0[\lambda (\phi) ,Q_0].
\end{equation}
Here  $E_0$ is the ground state energy of an equivalent single
junction with variable ratio $\lambda=E_J^{\rm tr}(\phi)/E_c \leq
2 \lambda_0$ and $q=Q_0$. $E_0$ is $2\pi$-periodic function of
$\phi$ and $2e$-periodic function of charge $Q_0$ and its shape
depends on the relation between $E_{J1,J2}$ and $E_c$.

Due to the periodic dependence on $\phi$, the Hamiltonian
Eq.~(\ref{Htr1}) can be expanded into a Fourier series
\begin{equation}
\label{Htr2} H_{\rm tr}(\phi,Q_0)= -\sum\limits_{k =
0}^\infty{E_J^{(k)}(Q_0 )\cos (k\phi )}.
\end{equation}
For $Q_0\neq e \bmod(2e)$ or/and at notable asymmetry of the
transistor, $E_{J1}\neq E_{J2}$, the terms with $k=0$ and 1 are
dominating in Eq.~(\ref{Htr2}). This is in accordance with almost
sinusoidal supercurrent-phase relation \cite{Zorin-IEEE} with
effective critical current $I_c^{\rm
tr}(Q_0)=\frac{2\pi}{\Phi_0}E_J^{(1)}(Q_0)$ modulated by gate
(see corresponding plots for different values of $\lambda_0$ in
Fig.~2a of Ref.\cite{Zorin-prl1}). In particular, in the symmetric
transistor with $\lambda_0\ll 1$, $I_c^{\rm tr}$ can be found
explicitly from Eqs.~(4)-(5) of Ref.~\cite{Zorin-prl1} using the
wave functions given by Eq.~(A11) of Ref.~\cite{LiZo}, i.e.,
\begin{equation}
\label{IcQ0}I_c^{\rm tr}(Q_0)= \frac{\lambda_0}{4}\left[1- \left(
Q_0/e \right)^2 \right]^{-1} I_c^{0}.
\end{equation}

In the degenerate case of $Q_0=e \bmod(2e)$ and $E_{J1}=
E_{J2}=E_J \lesssim E_c$, the ground state energy
Eq.~(\ref{Htr1}) is $E_0\approx -E_J |\cos\frac{\phi}{2}|+$const,
while the maximum critical current is $I_c^{\rm tr}(e)\approx
0.5I_{c1,c2}$. The Fourier coefficients in Eq.(\ref{Htr2}) are
equal to
\begin{equation}
\label{E-coeff}
E_J^{(k)}(e)=\frac{4}{\pi}\frac{(-1)^{k-1}}{(2k)^2-1}E_J,\qquad
k\geq 1,
\end{equation}
so the amplitudes of the principal (first harmonic) term
$E_J^{(1)}(e)\approx 0.42E_J$ and the ratio
$|E_J^{(2)}(e)/E_J^{(1)}(e)|=0.2.$

The expansion Eq.~(\ref{Htr2}) presents the transistor Hamiltonian
as a weighted sum of operators $e^{\pm ik\phi}$, each describing
tunneling of a cluster of $k$ Cooper pairs through the transistor
in the positive (negative) direction.  This presentation enables
particular rates for tunneling of $k$ pairs, $\Gamma_k(E)$, to be
found and the total current to be calculated via the corresponding
sum.

Due to finite impedance $R$, the phase $\phi$ couples to the
environmental degrees of freedom via an operator of phase on the
resistor $\phi_R$: $\phi = \frac{2\pi}{\Phi_0}Vt - \phi_R$. Phase
$\phi_R$ is related to the infinite set of bath variables. The
effect of this coupling can be treated by means of function $P$
determined by the charge of the tunneling particle (cluster of
particles), temperature and the real part of total impedance $Z$
as seen by the tunneling charge \cite{Devoret}.

In our case, the real part of the impedance is equal to
\begin{equation}
\label{ReZ} {\mathop{\rm Re}\nolimits} Z(\omega ) = {R
\mathord{\left/
 {\vphantom {R {\left[ {1 + \left( {\omega RC_{\rm tr} }
  \right)^2 } \right]}}} \right.
 \kern-\nulldelimiterspace} {\left[ {1 + \left( {\omega RC_{\rm tr} }
 \right)^2 } \right]}},
\end{equation}
where the transistor capacitance $C_{\rm tr}= \frac{C_1
C_2}{C_1+C_2}\approx \frac{C}{2}$ (corresponding to the
characteristic charging energy of $4E_c$) and the tunneling charge
is equal to $2ke$. The rates, $\Gamma_k (E)=\frac{\pi}{2\hbar}
[E_{J}^{(k)}]^2 P(E)$, and the total cotunneling current,
\begin{equation}
\label{Current-total} I_{\rm cot}=2e \sum\limits_{k =
1}^\infty{k\, [\Gamma_k(2keV)-\Gamma_k(-2keV)]},
\end{equation}
are found perturbatively, assuming $0.5E_J/4E_c=\lambda_0/8 \ll
1$. If $Q_0$ not close to $e$, the term with $k=1$ is essential,
and the cotunneling current is given by the formula similar to
that derived by Ingold et al. \cite{IngGrab} for tunneling across
a single junction at $2eV\ll \hbar/RC$,
\begin{eqnarray}
\label{I-V} I_{\rm cot} = &&\frac{\Phi_0 [I_c^{\rm
tr}(Q_0)]^2}{{4E_c }} \; e^{- 2\gamma \rho} \rho^{2\rho} \left(
{\frac{{2E_c }}{{\pi ^2 k_B T}}}
\right)^{1 - 2\rho}\nonumber\\
&& \times \frac{{\left| {\Gamma \left[ {\rho - j\left( {eV/\pi
k_B T} \right)} \right]} \right|^2 }}{{\Gamma (2\rho)}} \sinh
\left(\frac{eV}{k_BT} \right),
\end{eqnarray}
where $\rho = R/R_Q$, $\Gamma$ is Gamma function and
$\gamma\approx 0.577$ Euler's constant. The perturbation theory
result, Eqs.~(\ref{Current-total})-(\ref{I-V}), is valid if
$I_{\rm cot} \ll \rho I_c^{\rm tr}$. The latter condition means
that the cotunneling events are rare and incoherent and this is
nicely obeyed in our experiment.

\begin{figure}[t]
\begin{center}
\leavevmode
\includegraphics[width=2.5in]{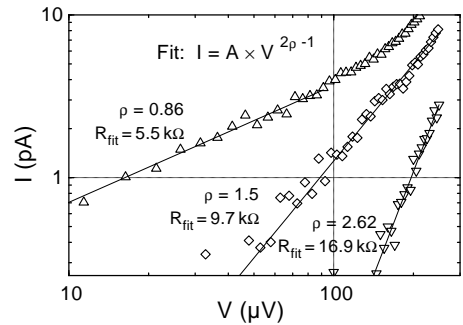}
\caption {The $I-V$ curves of three Cooper pair transistors with
different values of resistance $R$ (up-triangles - 5~k$\Omega$,
diamonds - 9~k$\Omega$ and down-triangles - 17~k$\Omega$, cf. the
fit values $R_{\rm{fit}}$) and their power-function fit in
accordance with Eq.~(\ref{low-V}). The fitted values of prefactor
$A=I_0 V_0^{1-2\rho}$ were used to restore the values of $E_J$
(=~33, 34 and 28~$\mu$eV, respectively).} \label{Scheme}
\end{center}
\end{figure}

Expression (\ref{I-V}) pretty well gives the transport and gate
voltage dependencies as well as the magnitude of the measured
current (see calculated curves in Fig.~1). The values of the
fitting parameters $R$ and $E_J$ are in good agreement with their
estimated values (10\% and 13\% deviation, respectively), while
the effective temperature $T_{\rm eff}$ of the resistor was found
to be 90~mK (cf. bath temperature of 20~mK). This value matches
with $T_{\rm eff}=70-80$~mK found in the same setup in
measurements of the three-junction Al-Cr single electron pump
with local resistors \cite{Pump} and can be ascribed mostly to a
noise effect.

We achieved the CTCP regime in the range of the gate charge
$|Q_0|\leq \frac{e}{2}$. The gate modulation depth of cotunneling
current was found to be $d\equiv I_{\rm cot}(\frac{e}{2})/I_{\rm
cot}(0)\approx 2$, that is in agreement with the value given by
Eq.~(\ref{IcQ0}), i.e. $d= [I_{c}^{\rm
tr}(\frac{e}{2})/I_{c}^{\rm tr}(0)]^2=\frac{16}{9} \approx 1.78$
\cite{factor2}. Approaching the points $Q_0=\pm e$ one could
expect an increase in the cotunneling current by a factor of
$(2/\lambda_0)^2 \approx 10^2$, but in these points STCP
dominates because effective dissipation for tunneling across
either junction is notably smaller than that for CTCP, $\rho
\rightarrow \rho/4$ \cite{Wilhelm}. As a manifestation of such
behavior in our samples, we observed a current increase by more
than two orders of magnitude when we applied a fast ($\sim~$1~MHz)
gate voltage sweep of amplitude $> e/C_g$. Such sweep allowed
quasiparticle poisoning of the island to be avoided for short
time intervals when passing the regions $Q_0\approx \pm e$.

Finally, an issue of particular interest is the suppression of
CTCP at larger dissipation, i.e. at $R>R_Q$ \cite{OBS-Zor}. The
increase in $R$ should lead to a blockade behavior of the $I-V$
curves at small $V$ and development of a broad current peak moving
to $V=\frac{4e}{C}$. The latter feature can hardly be clear
observed in experiment because at such a large voltage other
current components (quasiparticle and combined pair-quasiparticle
tunneling) are significant \cite{Kuzmin}. On the other hand, for
small voltages between $k_BT/2e$ and $e/C$, cotunneling still
dominates and for its estimation one can apply the zero
temperature expansion of function $P$ \cite{Devoret} and will get
\begin{equation}
\label{low-V} I_{\rm cot} = I_0
\left(\frac{V}{V_0}\right)^{2\rho-1},\,{\rm where} \,\;
I_0=\frac{\pi \rho e^{-2\gamma}\Phi_0 (I_c^{\rm tr})^2
}{32\Gamma(2\rho) E_c}
\end{equation}
and $V_0= \frac{2e}{\pi\rho C}$. Thus, the CTCP current obeys the
power law, i.e. it is efficiently suppressed at large $\rho$.
This effect was demonstrated and the results are presented in
Fig.~2. Although the experimental voltage range was rather
limited, the data are in good agreement with the dependence
Eq.~(\ref{low-V}) for parameter values close to those estimated
from measurements.

In conclusion, the obtained results on measurements of CTCP are
also of practical importance, e.g., for improving the performance
characteristics of the Cooper pair pumps. In the latter devices,
CTCP may significantly degrade the accuracy of charge pumping
\cite{Geerligs}. In contrast to the normal junction case, an
increase in the number $N$ of junctions in the superconducting
pump array seems to be not suitable for suppressing unwanted CTCP
\cite{Pekola}. On the other hand, the use of local resistors
\cite{Pump,CP-Rpump} ensures such suppression and, therefore, is
promising for the construction of an accurate Cooper pair pump
with a reasonable number ($N =3-5$) of junctions in the array.

This work was partially supported by the EU through the COUNT and
SQUBIT-2 projects.


\begin{references}
\bibitem{factor2} Note that formula for the
cotunneling rate presented in Ref.~\cite{Wilhelm}, i.e.
$\Gamma_{\rm CT}\propto (\delta E_{\rm ch})^{-2}$, gives the
overestimated value of $d=[\delta E_{\rm ch}(0)/\delta E_{\rm
ch}(\frac{e}{2})]^2=4$.
\end{references}
\end{document}